\documentclass[prd,reprint,showpacs,showkeys]{revtex4}
\usepackage{graphics}
\usepackage{eurosym}
\usepackage{amsfonts}
\usepackage{amssymb}
\usepackage{amsmath}
\usepackage{graphicx}
\usepackage[font={footnotesize,it}]{caption}
\usepackage{hyperref}

\makeatother

\begin{document}

\title{Hawking radiation of scalar and vector particles from 5D Myers-Perry
black holes}

\author{Kimet Jusufi}
\email{kimet.jusufi@unite.edu.mk}

\affiliation{Physics Department, State University of Tetovo, Ilinden Street nn,
1200, Macedonia}

\author{Ali \"{O}vg\"{u}n}
\email{ali.ovgun@emu.edu.tr}

\affiliation{Physics Department, Eastern Mediterranean University, Famagusta,
Northern Cyprus, Mersin 10, Turkey}

\date{\today }
\begin{abstract}
In the present paper we explore the Hawking radiation as a quantum
tunneling effect from a rotating 5 dimensional Myers-Perry black hole (5D-MPBH) with
two independent angular momentum components. First, we investigate the
Hawking temperature by considering the tunneling of massive scalar
particles and spin-1 vector particles from the 5D-MPBH in the Painlev\'{e} coordinates and then 
in the corotating frames. More specifically, we solve the Klein-Gordon and Proca equations by
applying the WKB method and Hamilton-Jacobi equation in both cases. Finally, we recover the Hawking 
temperature and show that coordinates systems do not affect the Hawking temperature.
\end{abstract}

\pacs{04.70.Dy, 04.62.+v, 11.30.Cp}

\keywords{5-D Myers-Perry Black Hole, Qunatum Tunneling, Klein-Gordon Equation,
Proca Equation }
\maketitle

\section{Introduction}

In 1974, Stephen Hawking suggests black holes cannot be completely black, and instead due to the quantum effects or quantum fluctuations they evaporate \cite{hawking}. Hawking discovered that if a quantum fluctuation takes at the event horizon, then, due to the quantum tunneling effect one of the virtual particles can be produced just inside the black hole, and the other particle outside the black hole. The particle produced inside the black hole will have a negative energy, this causes the black hole to lose mass and shrink over time and eventually after a very long period of time completely evaporate and disappear. On the other hand, the particle produced outside the black hole will have positive energy and can be detected by an observer as a real particle at infinity. 

After this amazing discovery, calculation of the Hawking radiation from different types of black holes with various methods become hot topics \cite{t1,t2,t3,t4,t5,t6,t7}. Recently, it has been speculated that the Hawking radiation and its entanglement in an analogue black hole has been observed \cite{stein}.
However, this result is debatable and needs more confirmation by other experiments. Furthermore, previously, the Hawking radiation of scalar bosons, spin-1 vector particles, spin-2 particles, spin-3/2 particles of different types of black holes and wormholes are studied \cite{majhi,t8,t9,t10,t11,t12,t13,t14,t15,t16,t17,t18,t19,t20,Kanti,sum1,sum2,canisius}. 

On the other hand, Hawking radiation from higher--dimensional black holes has attracted a lot of attention. For example, Hawking radiation from a rotating 5D-MPBH with two angular momentum components was investigated in Refs. \cite{frolov,zheng}, quantum anomalies in 5D-MPBH \cite{iso}, Hawking radiation of Dirac particles from a charged 5D-MPBH was investigated in \cite{ran,f1,f2,f3}, and recently the tunneling of Dirac particles under quantum gravity effects from 5D-MPBH with a single non-zero angular momentum was studied in \cite{5dgup}. However, vector particles  play a fundamental role in particle physics, for example, recently, there was an attempt to explain the origin of dark energy from a massive photon \cite{seyen}. In Refs. \cite{kahn,bednyakov} a massive photon or the so-called Darklight was studied to explain the origin of dark matter.

Inspired by this, in this paper, we calculate the Hawking temperature of scalar bosons
and spin-1 vector particles tunneling across the near horizon of the
black holes using the Hamilton-Jacobi method. In particular, we use the quantum tunneling method to study the Hawking radiation from a rotating 5D-MPBH with two angular momentum components of massive scalar/vector particles. Furthermore, we will aim to solve the problem by using different coordinates. Firstly, we will use the Painlev\'{e} coordinates and then we will eliminate the rotating degrees of freedom and introduce an appropriate coordinate transformation to the co-rotating frame \cite{zheng}.

The organization of this paper is as follows. In Section 2, we introduce
the rotating 5-dimensional Myers-Perry black hole (5D-MPBH) and in
Section 3, we investigate the tunneling of massive bosons and spin-1
particles from the 5D-MPBH in the Painlev\'{e} coordinates. The tunneling
of massive bosons and spin-1 particles from the 5D-MPBH in the corotating
frame is studied in Section 4. Section 5 is devoted to our conclusion.

\section{Rotating 5--dimensional Myers\textendash Perry black hole}

A rotating 5D-MPBH in $D=2n+1+\epsilon$
$(\epsilon=0\,\text{or}\,1)$ dimensions, with multiple nonzero angular
momentum parameters, can generally be written in the following form
\cite{frolov,zheng} 
\begin{equation}
\mathrm{d}s^{2}=-\mathrm{d}t^{2}+\epsilon r^{2}\mathrm{d}\alpha^{2}+\sum_{i=1}^{n}(r^{2}+a_{i}^{2})(\mathrm{d}\mu_{i}^{2}+\mu_{i}^{2}\mathrm{d}\phi_{i}^{2})+\frac{r_{0}^{2}r^{2-\epsilon}}{\Pi\mathcal{F}}\left(\mathrm{d}t-\sum_{i=1}^{n}a_{i}\mu_{i}^{2}\mathrm{d}\phi_{i}\right)^{2}+\frac{\Pi\mathcal{F}}{\Pi-r_{0}^{2}r^{2-\epsilon}}\mathrm{d}r^{2}
\end{equation}
where
\begin{equation}
\mathcal{F}=1-\sum_{i=1}^{n}\frac{a_{i}^{2}\mu_{i}^{2}}{r^{2}+a_{i}^{2}},
\end{equation}
and 
\begin{equation}
\Pi=\prod_{i=1}^{n}(r^{2}+a_{i}^{2}).
\end{equation}

Moreover $\mu_{i}$ and $\alpha$ are related as
\begin{equation}
\sum_{i=1}^{n}\mu_{i}^{2}+\epsilon\alpha^{2}=1,\,\,\,\,(0\leq\mu_{i}\leq1),\,\,\,\,(-1\leq\alpha\leq1).
\end{equation}

A rotating black hole in $D=5$, rotates in each $\phi_{i}$\textendash direction,
so there are $(D-1)/2$ angular momentum components. Next, let us introduce
$\mu_{1}=\cos\theta$, $\mu_{2}=\sin\theta$, $\phi_{1}=\phi$, and $\phi_{2}=\psi$; then we end up with the
following metric:
\begin{eqnarray}\label{5}
\mathrm{d}s^{2} & = & g_{\mu\nu}\mathrm{d}x^{\mu}\mathrm{d}x^{\nu}\\
 & = & -\mathrm{d}t^{2}+\frac{\rho^{2}r^{2}}{\Delta}\mathrm{d}r^{2}+\rho^{2}\mathrm{d}\theta^{2}+(r^{2}+a^{2})\sin^{2}\theta\,\mathrm{d}\phi^{2}+(r^{2}+b^{2})\cos^{2}\theta\,\mathrm{d}\psi^{2}+\frac{r_{0}^{2}}{\rho^{2}}\left(\mathrm{d}t-a\sin^{2}\theta\mathrm{d}\phi-b\cos^{2}\theta\,\mathrm{d}\psi\right)^{2},\nonumber  
\end{eqnarray}
where 
\begin{equation}
\rho^{2}=r^{2}+a^{2}\cos^{2}\theta+b^{2}\sin^{2}\theta,\,\,\,\,\,\,\Delta=(r^{2}+a^{2})(r^{2}+b^{2})-r_{0}^{2}r^{2},\,\,a_{1}=a,\,\,\,a_{2}=b.
\end{equation}

Note that $r_{0}$ is length parameter connected with the black hole
mass $M$ by
\begin{equation}
M=\frac{3r_{0}^{2}}{8\sqrt{\pi}G},
\end{equation}
where $G$ is $(4+1)$ dimensional gravitational constant. On the other
hand the parameters $a$ and $b$ are associated with its two independent
angular momenta, respectively.

Solving $g^{rr}(r_{+})=0$ one can find the relation for the event horizon given by 
\begin{equation}
r_{\pm}^{2}=\frac{1}{2}\left[r_{0}^{2}-a^{2}-b^{2}\pm\sqrt{(r_{0}^{2}-a^{2}-b^{2})^{2}-4a^{2}b^{2}}\right].
\end{equation}

We can now choose a new coordinate frame, co--rotating with the black hole
horizon, to eliminate the dragging motion on the rotating degrees of
freedom of a tunneling particle by using the following coordinate
transformations 
\begin{equation}
\mathrm{d}\phi=\mathrm{d}\tilde{\phi}+\Omega_{a}\mathrm{d}t,\,\,\,\,\mathrm{d}\psi=\mathrm{d}\tilde{\psi}+\Omega_{b}\mathrm{d}t,
\end{equation}
in which the corresponding angular velocities at the horizon are given by 
\begin{equation}
\Omega_{a}=\frac{a}{r^{2}+a^{2}},\,\,\,\,\Omega_{b}=\frac{b}{r^{2}+b^{2}}.
\end{equation}

Then the metric \eqref{5} reads 
\begin{eqnarray}\label{11}
\mathrm{d}s^{2} & = & -G_{tt}(r,\theta,\phi,\psi)\,\mathrm{d}t^{2}+\frac{r^{2}\rho^{2}}{\Delta}\mathrm{d}r^{2}+\rho^{2}\mathrm{d}\theta^{2}+\left[(r^{2}+a^{2})+\frac{r_{0}^{2}a^{2}\sin^{2}\theta}{\rho^{2}}\right]\sin^{2}\theta\,\mathrm{d}\tilde{\phi}^{2}\nonumber \\
 & + & \left[(r^{2}+b^{2})+\frac{r_{0}^{2}b^{2}\cos^{2}\theta}{\rho^{2}}\right]\cos^{2}\theta\,\mathrm{d}\tilde{\psi}^{2}+\frac{2\,a\,b\,r_{0}^{2}}{\rho^{2}}\sin^{2}\theta\cos^{2}\theta\,\mathrm{d}\tilde{\phi}\,\mathrm{d}\tilde{\psi}
\end{eqnarray}
where 
\begin{equation}
G_{tt}(r,\theta,\phi,\psi)=g_{tt}-g_{\phi\phi}\Omega_{a}^{2}-g_{\psi\psi}\Omega_{b}^{2}+2g_{t\phi}\Omega_{a}+2g_{t\psi}\Omega_{b}-2g_{\phi\psi}\Omega_{a}\Omega_{b}. \label{12}
\end{equation}

Note that in the coordinate frame which is co--rotating with the event horizon $g_{t \tilde{\phi}_{i}}=g_{t \phi_{i}}-g_{\phi_{i} \phi_{j}}\Omega_{j}$, should be zero at the horizon. i.e., $g_{t \tilde{\phi}_{i}}(r_{+})=0$ \cite{zheng}. By taking this into account Eq. \eqref{12} simplifies to 
\begin{equation}
G_{tt}(r_{+})= g_{tt}+g_{t\phi}\Omega_{a}+g_{t\psi}\Omega_{b},
\end{equation}
in which 
\begin{eqnarray}
g_{rr} & = & \frac{r^{2}\rho^{2}(r)}{\Delta(r)},\\
g_{tt} & = & 1-\frac{r_{0}^{2}}{\rho^{2}},\\
g_{t\phi} & = & \frac{ar_{0}^{2}\sin^{2}\theta}{\rho^{2}},\\
g_{t\psi} & = & \frac{br_{0}^{2}\cos^{2}\theta}{\rho^{2}}.\\
\end{eqnarray}

The Hawking temperature of the rotating 5-D MPBH in the units $k_{B}=c=G=\hbar=1$ can be computed as follows \cite{frolov,zheng}
\begin{equation}\label{19}
T_{H}=\frac{\kappa(r_{+})}{2\pi}=\lim_{r\to r_{+}}\frac{\partial_{r}\sqrt{G_{tt}}}{2\pi\sqrt{g_{rr}}}=\left.\frac{\partial_{r}\Pi-2r_{0}^{2}r}{4\pi r_{0}^{2}r^{2}}\right\vert _{r=r_{+}}
\end{equation}

\section{Quantum tunneling in Painlev\'{e} coordinates}

\subsection{Tunneling of vector particles}

The particle being emitted by the black hole should not depend on
some fixed azimuthal angles $(\theta_{0},\phi_{0},\psi_{0})$. If we introduce
the following Painlev\'{e} coordinate transformation into the Eq. \eqref{11}
\begin{equation}\label{20}
\mathrm{d}t=\mathrm{d}T-\sqrt{\frac{g_{rr}(r,\theta_{0},\phi_{0},\psi_{0})-1}{G_{tt}(r,\theta_{0},\phi_{0},\psi_{0})}}\mathrm{d}r,
\end{equation}
then keeping in mind that for a fixed angles in the co--rotating frame the tunneling
particles should satisfy $\mathrm{d}\tilde{\phi}=\mathrm{d}\tilde{\psi}=0$, we find the following
 metric \cite{zheng} 
\begin{equation}\label{21}
\mathrm{d}s^{2}=-F(r)\mathrm{d}T^{2}+2\sqrt{F(r)}\sqrt{H(r)-1}\,\mathrm{d}r\mathrm{d}T+\mathrm{d}r^{2}
\end{equation}
in which  $T$ is the Painlev\'{e} coordinate time and
\begin{equation}
F(r)=G_{tt}(r),\,\,\,\,H(r)=g_{rr}(r).
\end{equation}

The motion of a massive vector particle, described by the vector field
$\Psi^{\mu}$, can be studied by the Proca equation (PE), which reads \cite{t11}
\begin{eqnarray}\label{23}
\frac{1}{\sqrt{-G}}\partial_{\mu}\left(\sqrt{-G}\,\Psi^{\mu\nu}\right)-\frac{m^{2}}{\hbar^{2}}\Psi^{\nu}=0,
\end{eqnarray}
where $G=\det G_{\mu\nu}=FH$ and 
\begin{equation}
\Psi_{\mu\nu}=\partial_{\mu}\Psi_{\nu}-\partial_{\nu}\Psi_{\mu}.
\end{equation}

The PE equation in the spacetime metric \eqref{21} can be solved by applying the WKB approximation method which suggest that 
\begin{equation}\label{25}
\Psi_{\nu}=C_{\nu}(T,r)\exp\left(\frac{i}{\hbar}\left(S_{0}(T,r)+\hbar\,S_{1}(T,r)+\hdots.\right)\right).
\end{equation}

Furthermore by considering the spacetime symmetries of the metric
\eqref{21}, the following ansatz for the action can be chosen 
\begin{equation}\label{26}
S_{0}(T,r)=-E\,T+R(r),
\end{equation}
in which $E$ is the energy of the particle. We can now insert Eq. \eqref{25} into the Eq. \eqref{23} and keep only the leading order of $\hbar$. Hereinafter, we have two differential equations:

\begin{eqnarray}
\frac{\left(-\sqrt{F}m^{2}\sqrt{H-1}+ER^{\prime}\right)C_{1}}{FH}+\frac{\left((R^{\prime})^{2}+m^{2}\right)C_{2}}{FH} & = & 0,\\
\frac{\left(-Fm^{2}+E^{2}\right)C_{1}}{FH}+\frac{\left[E\sqrt{F}R^{\prime}\sqrt{H-1}-\sqrt{F}m^{2}(H-1)\right]C_{2}}{FH\sqrt{H-1}} & = & 0
\end{eqnarray}

With the non-zero elements of the matrix $\mathbb{M}$: 
\begin{eqnarray}
\mathbb{M}_{11} & = & \frac{-\sqrt{F}m^{2}\sqrt{H-1}+ER^{\prime}}{FH}\\
\mathbb{M}_{12} & = & \frac{(R^{\prime})^{2}+m^{2}}{FH},\\
\mathbb{M}_{21} & = & \frac{-Fm^{2}+E^{2}}{FH},\\
\mathbb{M}_{22} & = & \frac{E\sqrt{F}R^{\prime}\sqrt{H-1}-\sqrt{F}m^{2}(H-1)}{FH\sqrt{H-1}}
\end{eqnarray}

Solving the determinant 
\begin{equation}
\det \mathbb{M}(C_{1},C_{2})^{T}=0,
\end{equation}
we find the following equation 
\begin{equation}
-\frac{m^{2}\left(F^{3/2}H\sqrt{H-1}m^{2}-F^{3/2}\sqrt{H-1}(R^{\prime})^{2}+E^{2}\sqrt{F}\sqrt{H-1}+2EFR^{\prime}(H-1)\right)}{H^{2}F^{3/2}\sqrt{H-1}}=0.
\end{equation}

Let us now solve this equation for the radial trajectories to get the following
integral 
\begin{equation}\label{35}
R_{\pm}(r)=\int\left(\frac{E\sqrt{H-1}}{\sqrt{F}}\pm\frac{\sqrt{H}\sqrt{E^{2}-m^{2}F}}{\sqrt{F}}\right) \mathrm{d}r.
\end{equation}

In order to calculate the tunneling rate, one faces the well known factor--two problem (see for example \cite{Akhmedova1,borun}).
The right way to solve this problem is to consider first the invariance under canonical transformations given as  $\oint p_{r}\mathrm{d}r=\int p_{r}^{+}\mathrm{d}r-\int p_{r}^{-}\mathrm{d}r$. We can first calculate the spatial contribution of the imaginary part of $\text{Im}\, R(r)$. To do so, first we note that there is a pole at the horizon $r = r_{+}$ of the Eq. \eqref{35}, since $F(r_{+}) = 0$ and $\mathcal{H}(r_{+})=0$, where we have used the relation $\mathcal{H}=H^{-1}$. Thus if we shift the pole into the upper half plane $r_{+}\to r_{+}+i\epsilon$ and take the imaginary part of Eq. \eqref{35} we find 
\begin{equation}
\text{Im} \oint p_{r} \mathrm{d}r= \lim_{\epsilon\to 0}\left\lbrace\text{Im}\oint \frac{E\sqrt{1-\mathcal{H}}\pm\sqrt{E^{2}-m^{2}F}}{\sqrt{\mathcal{H}^{\prime}(r_{+},\theta_{0},\phi_{0},\psi_{0}) \,F^{\prime}(r_{+},\theta_{0},\phi_{0},\psi_{0})}(r-r_{+}\pm i\epsilon)}\mathrm{d}r\right\rbrace,
\end{equation}
where we have used the relation $p_{r}=\partial_{r}R$.  Furthermore the above equation can also be written as
\begin{eqnarray}
\text{Im} \oint p_{r} \mathrm{d}r=\lim_{\epsilon \to 0}\left\lbrace\text{Im}\left[\int_{r_{i}}^{r_{f}} \frac{E\sqrt{1-\mathcal{H}}+\sqrt{E^{2}-m^{2}F}}{\sqrt{\mathcal{H}^{\prime}(r_{+}) F^{\prime}(r_{+})}(r-r_{+}+ i\epsilon)} \mathrm{d}r+\int_{r_{f}}^{r_{i}} \frac{E\sqrt{1-\mathcal{H}}-\sqrt{E^{2}-m^{2}F}}{\sqrt{\mathcal{H}^{\prime}(r_{+}) F^{\prime}(r_{+})}(r-r_{+}- i\epsilon)}  \mathrm{d}r \right]\right\rbrace.
\end{eqnarray}

The physical meaning of this integral is that it gives the total spatial contribution and that's why we are calculating for a round trip. One can immediately observe from the last equation that there is no contribution to the imaginary part from the second term. However this is  not a surprising result since we are using Painlev\'{e} coordinates and we know that the particle experiences barrier only from inside the horizon to outside and not the other way. Now we make use of the equation
\begin{equation}\label{38}
\lim_{\epsilon \to 0}\text{Im} \frac{1}{r-r_{+}\pm i\epsilon}=\pi \delta (r-r_{+}).
\end{equation}

For the imaginary part of the first term we find
\begin{equation}
\text{Im} \oint p_{r} \mathrm{d}r=\frac{2\pi E }{\sqrt{\mathcal{H}^{\prime}(r_{+}) F^{\prime}(r_{+})}}.
\end{equation}

Now we have to calculate the temporal contribution.  From Eq. \eqref{20} the Painlev\'{e} coordinate time reads
\begin{equation}
T=t+\int \frac{\sqrt{1-\mathcal{H}(r,\theta_{0},\phi_{0},\psi_{0})}}{\sqrt{\mathcal{H}(r,\theta_{0},\phi_{0},\psi_{0}) \,F(r,\theta_{0},\phi_{0},\psi_{0})}} \mathrm{d}r.
\end{equation}

Substituting this result into the action \eqref{26} gives
\begin{equation}
S_{0}(T,r)=-Et-E\int \frac{\sqrt{1-\mathcal{H}(r,\theta_{0},\phi_{0},\psi_{0})}}{\sqrt{\mathcal{H}(r,\theta_{0},\phi_{0},\psi_{0}) \,F(r,\theta_{0},\phi_{0},\psi_{0})}} \mathrm{d}r+R(r).
\end{equation}

Therefore for the temporal contribution we find
\begin{equation}
\text{Im}(E\Delta T^{out,in})=-\frac{\pi E}{\sqrt{\mathcal{H}^{\prime}(r_{+}) F^{\prime}(r_{+})}}.
\end{equation}

According to Akhmedova et al \cite{Akhmedova1}, we can find the resulting tunneling rate by putting all these results together
\begin{eqnarray}
\Gamma &=&\exp \left[\frac{1}{\hbar}\left(\text{Im} (E\Delta T^{out})+\text{Im}(E\Delta T^{in})-\text{Im }\oint p_{r} \mathrm{d}r\right)\right]\\
&=& \exp \left(-\frac{4\pi E }{\sqrt{\mathcal{H}^{\prime}(r_{+}) F^{\prime}(r_{+})}}\right).
\end{eqnarray}

And finally we can find the Hawking temperature by comparing the latter
result with the Boltzmann formula $\Gamma_{B}=e^{-E/T_{H}}$ and setting
the Planck constant to unity to get
\begin{equation}
T_{H}=\frac{\sqrt{ \mathcal{H}^{\prime}(r_{+}) F^{\prime}(r_{+})}}{4 \pi}
\end{equation}

On the other hand, if we consider the expansions of $\mathcal{H}(r)$ and $F(r)$ in Taylor's series near the horizon given as \cite{zheng}
\begin{eqnarray}
\mathcal{H}(r) & = & \left.\frac{\partial_{r}\Pi-2r_{0}^{2}r}{r^{2}\rho^{2}}\right\vert _{r=r_{+}}(r-r_{+})+\hdots\\
F(r) & = & \left.\frac{(\partial_{r}\Pi-2r_{0}^{2}r)\,r^{2}\rho^{2}}{r_{0}^{4}r^{4}}\right\vert _{r=r_{+}}(r-r_{+})+\hdots
\end{eqnarray}

We recover the correct Hawking temperature for the 5D-MPBH given by 
\begin{equation}
T_{H}=\frac{\partial_{r}\Pi-2r_{0}^{2}r_{+}}{4\pi r_{0}^{2}r_{+}^{2}}.
\end{equation}

As expected, this result is in agreement with Eq. \eqref{19}.

\bigskip
\bigskip

\subsection{Tunneling of scalar particles}

Let us now consider the Klein-Gordon equation in curved spacetime
metric \eqref{21} for a massive scalar field $\Phi$ given as follows

\begin{equation}
\frac{1}{\sqrt{-G}}\partial_{\mu}\left(\sqrt{-G}\,G^{\mu\nu}\partial_{\nu}\Phi\right)-\frac{m^{2}}{\hbar^{2}}\Phi=0.\label{48}
\end{equation}

In which we have used $G=\det G_{\mu\nu}$ and $m$ is the mass of
the scalar particle. This equation can be solved by using the semiclassical
WKB approximation which allows us to choose the following ansatz for
the scalar field: 
\begin{equation}
\Phi(T,r)=\exp{\left(\frac{i}{\hbar}S\left(T,r\right)\right)},\label{49}
\end{equation}
where $S(T,r)$ is the classically forbidden action for the tunneling.
Inserting the above scalar field $\Phi$ into the Eq. \eqref{48} we end up with the
following expression:

\begin{equation}
\frac{1}{FH}(\partial_{T}S_{0})^{2}=\frac{1}{H}(\partial_{r}S_{0})^{2}+2\frac{\sqrt{H-1}}{\sqrt{F}H}(\partial_{r}S_{0})(\partial_{T}S_{0})+m^{2}.
\end{equation}

Choosing the action as \eqref{26} and solving for the radial part we
get 
\begin{equation}
\frac{1}{FH}E^{2}=\frac{1}{H}(\partial_{r}R)^{2}-2E\frac{\sqrt{H-1}}{\sqrt{F}H}(\partial_{r}R)+m^{2}.
\end{equation}

From where one can obtain the same result for the radial part as in the last section 
\begin{equation}
R_{\pm}(r)=\int\left(\frac{E\sqrt{H-1}}{\sqrt{F}}\pm\frac{\sqrt{H}\sqrt{E^{2}-m^{2}F}}{\sqrt{F}}\right)\mathrm{d}r.
\end{equation}

And we recover the same Hawking temperature of scalar particles
\begin{equation}
T_{H}=\frac{\partial_{r}\Pi-2r_{0}^{2}r_{+}}{4\pi r_{0}^{2}r_{+}^{2}}.
\end{equation}

\section{Quantum tunneling in the corotating frame}

\subsection{Tunneling of vector particles}

We have shown in the first section that we can eliminate the frame dragging effects on the tunneling particle by introducing the corotating
frame. If we drop the tilda notation in the metric \eqref{11} we find  
\begin{equation}\label{54}
\mathrm{d}s^{2}=-F(r_{+},\theta)\mathrm{d}t^{2}+H(r_{+},\theta)\mathrm{d}r^{2}+K(r_{+},\theta)\mathrm{d}\theta^{2}+M(r_{+},\theta)\mathrm{d}\phi^{2}+N(r_{+},\theta)\mathrm{d}\psi^{2}+2P(r_{+},\theta)\mathrm{d}\phi\,\mathrm{d}\psi,
\end{equation}
in which 
\begin{eqnarray}
F(r_{+},\theta) & = & g_{tt}+g_{t\phi}\Omega_{a}+g_{t\psi}\Omega_{b},\\
H(r_{+},\theta) & = & \frac{r_{+}^{2}\rho^{2}(r_{+})}{\Delta(r_{+})},\\
K(r_{+},\theta) & = & \rho^{2}(r_{+}),\\
M(r_{+},\theta) & = & \left[(r_{+}^{2}+a^{2})+\frac{r_{0}^{2}a^{2}\sin^{2}\theta}{\rho^{2}}\right]\sin^{2}\theta,\\
N(r_{+},\theta) & = & \left[(r_{+}^{2}+b^{2})+\frac{r_{0}^{2}b^{2}\cos^{2}\theta}{\rho^{2}}\right]\cos^{2}\theta,\\
P(r_{+},\theta) & = & \frac{a\,b\,r_{0}^{2}}{\rho^{2}}\sin^{2}\theta\cos^{2}\theta,
\end{eqnarray}

Let us now recall again that the PE of the vector field
$\Psi^{\mu}$, in the spacetime metric \eqref{54} reads
\begin{eqnarray}\label{61}
\frac{1}{\sqrt{-G}}\partial_{\mu}\left(\sqrt{-G}\,\Psi^{\mu\nu}\right)-\frac{m^{2}}{\hbar^{2}}\Psi^{\nu}=0,
\end{eqnarray}
where this time $G=\det G_{\mu\nu}=HKF(MN-P^{2})$. We apply the same method, namely, we consider the WKB approximation method 
\begin{equation}\label{62}
\Psi_{\nu}=C_{\nu}(t,r,\theta,\phi,\psi)\exp\left(\frac{i}{\hbar}\left(S_{0}(t,r,\theta,\phi,\psi)+\hbar\,S_{1}(t,r,\theta,\phi,\psi)+\hdots.\right)\right).
\end{equation}

Taking into the consideration the symmetries of the metric \eqref{54} given
by three corresponding Killing vectors $(\partial/\partial_{t})^{\mu}$,
$\partial/\partial_{\phi})^{\mu}$ and $(\partial/\partial_{\psi})^{\mu}$,
we may choose the following ansatz for the action 
\begin{equation}
S_{0}(t,r,\theta,\phi,\psi)=-(E-(j\Omega_{a}+l\Omega_{b}))t+R(r,\theta)+j\phi+l\psi,
\end{equation}
in which $E$ is the energy of the particle, and $j$ and $l$ are the angular momentum of the particle corresponding
to the angles $\phi$ and $\psi$, respectively. If we now insert
the Eq. \eqref{62} into the Eq. \eqref{61} and keep only the leading order of
$\hbar$ we find the following set of five differential equations:
\bigskip
\begin{eqnarray}
0 & = & \frac{\tilde{E}R^{\prime}(r)C_{1}}{FH}+\frac{\tilde{E}(\partial_{\theta}R(r))C_{2}}{FK}+\frac{\left[(\Omega_{a}lj-l(E-l\Omega_{b}))P+\tilde{E}jN\right]C_{3}}{F\zeta}+\frac{\left[(\Omega_{a}j^{2}+(l\Omega_{b}-E)j)P+l\tilde{E}M\right]C_{4}}{F\zeta}\nonumber \\
 & + & \frac{\left[H\zeta(\partial_{\theta}R)^{2}+K\left(\zeta(R^{\prime})^{2}+H(-m^{2}P^{2}-2jlP+(Nm^{2}+l^{2})M+j^{2}N)\right)\right]C_{5}}{FH\zeta K},\\[0.5cm]
0 & = & \frac{\left[-F\zeta(\partial_{\theta}R)^{2}+\left((m^{2}F-\tilde{E}^{2})P^{2}+2jlFP+((-m^{2}F+\tilde{E}^{2})N-Fl^{2})M-FNj^{2}\right)K\right]C_{1}}{FH\zeta K}+\frac{(\partial_{\theta}R)R^{\prime}C_{2}}{HK}\nonumber \\
 & + & \frac{(Nj-lP)R^{\prime}(r)C_{3}}{H\zeta}+\frac{(lM-jP)R^{\prime}(r)C_{4}}{H\zeta}+\frac{\tilde{E}R^{\prime}C_{5}}{FH},\\[0.5cm]
0 & = & \frac{R^{\prime}(\partial_{\theta}R)C_{1}}{HK}+\frac{\left[-F\zeta(\partial_{\theta}R)^{2}+\left((m^{2}F-\tilde{E}^{2})P^{2}+2jlFP+((-m^{2}F+\tilde{E})N-Fl^{2})M-FNj^{2}\right)H\right]C_{2}}{HF\zeta K}\nonumber \\
 & + & \frac{(Nj-lP)(\partial_{\theta}R)C_{3}}{\zeta K}+\frac{(lM-jP)(\partial_{\theta}R)C_{4}}{\zeta K}+\frac{\tilde{E}(\partial_{\theta}R)C_{5}}{FK},\\[0.5cm]
0 & = & \frac{(Nj-lP)R^{\prime}C_{1}}{\zeta H}+\frac{(Nj-lP)(\partial_{\theta}R)C_{2}}{\zeta K}+\frac{\left[-FHN(\partial_{\theta}R)^{2}+\left(-FN(R^{\prime})^{2}+H((-m^{2}F+\tilde{E}^{2})N-Fl^{2})\right)K\right]C_{3}}{\zeta HKF}\nonumber \\
 & + & \frac{\left[FHP(\partial_{\theta}R)^{2}+\left(FP(R^{\prime})^{2}+H((m^{2}F-\tilde{E}^{2})P+Fjl)\right)K\right]C_{4}}{\zeta HKF}+\frac{\tilde{E}(jN-lP)C_{5}}{\zeta F},\\[0.5cm]
0 & = & \frac{(Ml-jP)(R^{\prime})C_{1}}{\zeta H}+\frac{(Ml-jP)(\partial_{\theta}R)C_{2}}{\zeta K}+\frac{\left[FHP(\partial_{\theta}R)^{2}+\left(FP(R^{\prime})^{2}+H((m^{2}F-\tilde{E}^{2})P+Flj)\right)K\right]C_{3}}{\zeta HKF}\nonumber \\
 & + & \frac{\left[-FHM(\partial_{\theta}R)^{2}+\left(-FM(R^{\prime})^{2}+H((-m^{2}F+\tilde{E}^{2})M-Fj^{2})\right)K\right]C_{4}}{\zeta HKF}+\frac{\tilde{E}(lM-jP)C_{5}}{\zeta F},
\end{eqnarray}

in which $\zeta=MN-P^{2}$ and $\tilde{E}=E-(j\Omega_{a}+l\Omega_{b})$.
From this set of five equations we can construct a $5\times5$
matrix $\aleph$, which satisfies the following matrix equation 
\begin{equation}
 \aleph(C_{1},C_{2},C_{3},C_{4},C_{5})^{T}=0.
\end{equation}

Using the last equation we find the following non\textendash zero matrix elements: 
\begin{eqnarray*}
\aleph_{11} & = & \aleph_{25}=\frac{\tilde{E}R^{\prime}(r)}{FH},\\
\aleph_{12} & = & \aleph_{35}=\frac{\tilde{E}(\partial_{\theta}R(r))}{FK},\\
\aleph_{13} & = & \frac{(\Omega_{a}lj-l(E-l\Omega_{b}))P+\tilde{E}jN}{F\zeta},\\
\aleph_{14} & = & \frac{(\Omega_{a}j^{2}+(l\Omega_{b}-E)j)P+\tilde{E}lM}{F\zeta},\\
\aleph_{15} & = & \frac{H\zeta(\partial_{\theta}R)^{2}+K\left(\zeta(R^{\prime})^{2}+H(-m^{2}P^{2}-2jlP+(Nm^{2}+l^{2})M+j^{2}N)\right)}{FH\zeta K},\\
\aleph_{21} & = & \frac{-F\zeta(\partial_{\theta}R)^{2}+\left((m^{2}F-\tilde{E}^{2})P^{2}+2jlFP+((-m^{2}F+\tilde{E}^{2})N-Fl^{2})M-FNj^{2}\right)K}{FH\zeta K},\\
\aleph_{22} & = & \aleph_{31}=\frac{(\partial_{\theta}R)R^{\prime}}{HK},\\
\aleph_{23} & = & \aleph_{41}=\frac{(Nj-lP)R^{\prime}(r)}{H\zeta},\\
\aleph_{24} & = & \aleph_{51}=\frac{(lM-jP)R^{\prime}(r)}{H\zeta},\\
\aleph_{32} & = & \frac{-F\zeta(\partial_{\theta}R)^{2}+\left((m^{2}F-\tilde{E}^{2})P^{2}+2jlFP+((-m^{2}F+\tilde{E})N-Fl^{2})M-FNj^{2}\right)H}{HF\zeta K},\\
\aleph_{33} & = & \aleph_{42}=\frac{(Nj-lP)(\partial_{\theta}R)}{\zeta K},\\
\aleph_{34} & = & \aleph_{52}=\frac{(lM-jP)(\partial_{\theta}R)}{\zeta K},\\
\aleph_{43} & = & \frac{-FHN(\partial_{\theta}R)^{2}+\left(-FN(R^{\prime})^{2}+H((-m^{2}F+\tilde{E}^{2})N-Fl^{2})\right)K}{\zeta HKF},\\
\aleph_{44} & = & \aleph_{53}=\frac{FHP(\partial_{\theta}R)^{2}+\left(FP(R^{\prime})^{2}+H((m^{2}F-\tilde{E}^{2})P+Fjl)\right)K}{\zeta HKF},\\
\aleph_{45} & = & \frac{\tilde{E}(jN-lP)}{\zeta F}\\
\aleph_{54} & = & \frac{-FHM(\partial_{\theta}R)^{2}+\left(-FM(R^{\prime})^{2}+H((-m^{2}F+\tilde{E}^{2})M-Fj^{2})\right)K}{\zeta HKF}\\
\aleph_{25} & = & \frac{\tilde{E}(lM-jP)}{\zeta F}.
\end{eqnarray*}

If we solve $\det \aleph =0$, we get the following result
\begin{equation}
\frac{\left[-HF\zeta(\partial_{\theta}R)^{2}+K\left(-F\zeta(R^{\prime})^{2}+H(((-m^{2}F+\tilde{E})N-Fl^{2})M-FNj^{2}-P((-m^{2}F+\tilde{E})P-2Fjl))\right)\right]^{4}m^{2}}{H^{5}F^{5}\zeta^{5}K^{5}}=0.
\end{equation}

We solve for the radial part to get the following integral 
\begin{equation}
R(r)=\pm\int \frac{1}{\sqrt{\mathcal{H}^{\prime}(r_{+}) F^{\prime}(r_{+})}}\sqrt{(E-(j\Omega_{a}+l\Omega_{b}))^{2}-F(r)\left[m^{2}+\frac{1}{MN-P^{2}}\left(Nj^{2}+Ml^{2}-2Pjl\right)+\frac{(\partial_{\theta}R)^{2}}{K}\right]}\mathrm{d}r.
\end{equation}

\bigskip

We can now calculate the Hawking temperature. To do so we can fix the angle $\theta=\theta_{0}$, after that as we know we have to carry out first the spatial part contribution to the imaginary part of $\text{Im} \,R(r)$ and then the temporal part contribution.  The spatial part contribution can be easily found if we solve the last integral using \eqref{38} which leads us to the following result
\begin{equation}
\text{Im}R_{\pm}(r)=\pm \frac{\pi \tilde{E}}{\sqrt{{\mathcal{H}^{\prime}(r_{+},\theta_{0},\phi_{0},\psi_{0}) F^{\prime}(r_{+},\theta_{0},\phi_{0},\psi_{0})}}}.
\end{equation}

Therefore the spatial contribution to the tunneling rate gives
\begin{eqnarray}
\exp\left(-\frac{1}{\hbar}\text{Im} \oint p_{r} \mathrm{d}r\right)&=&\exp\left[-\frac{1}{\hbar} \text{Im} \left(\int p_{r}^{+}\mathrm{d}r-\int p_{r}^{-}\mathrm{d}r\right) \right]\\
&=& \exp \left(- \frac{2 \pi \tilde{E}}{\sqrt{{\mathcal{H}^{\prime}(r_{+},\theta_{0},\phi_{0},\psi_{0}) F^{\prime}(r_{+},\theta_{0},\phi_{0},\psi_{0})}}}\right).
\end{eqnarray}

The temporal part contribution comes due to the connection of the interior region and the exterior region of
the black hole. Thus, if one introduces $t\to t -i\pi/(2\kappa) $, one will have Im ($\tilde{E}\Delta t^{out,in})=-\tilde{E}\pi/(2\kappa)$. Then the total temporal contribution for a round trip can be calculated as
\begin{equation}\label{75}
\text{Im}(\tilde{E}\Delta t^{out,in})=-\frac{\pi \tilde{E}}{\kappa},
\end{equation}
where the surface gravity is given as follows \cite{zheng}
\begin{equation}
\kappa =\lim_{r\to r_{+}} \frac{\partial_{r}\sqrt{F(r)}}{\sqrt{H(r)}}.
\end{equation}

Then Eq. \eqref{75} takes the form 
\begin{equation}
\text{Im}(\tilde{E}\Delta t^{out,in})=-\frac{2 \pi \tilde{E}}{\sqrt{{\mathcal{H}^{\prime}(r_{+},\theta_{0},\phi_{0},\psi_{0}) F^{\prime}(r_{+},\theta_{0},\phi_{0},\psi_{0})}}}.
\end{equation}

The resulting tunneling rate is calculated as
\begin{eqnarray}
\Gamma & = & \exp\left[\frac{1}{\hbar}\left(\text{Im}(\tilde{E}\Delta t^{out})+ \text{Im}(\tilde{E}\Delta t^{in})-\text{Im}\oint p_{r} \mathrm{d}r \right)\right]\\
&=& \exp \left[- \frac{4 \pi \tilde{E}}{\sqrt{{\mathcal{H}^{\prime}(r_{+},\theta_{0},\phi_{0},\psi_{0}) F^{\prime}(r_{+},\theta_{0},\phi_{0},\psi_{0})}}}\right].
\end{eqnarray}

Comparing this result with the Boltzmann factor $e^{-\tilde{E}/T_{H}}$, one gets the correct Hawking temperature \eqref{19}
\begin{equation}
T_{H}=\frac{\partial_{r}\Pi-2r_{0}^{2}r_{+}}{4\pi r_{0}^{2}r_{+}^{2}}.
\end{equation}

\subsection{Tunneling of scalar particles}

The Klein-Gordon equation in curved spacetime metric \eqref{54} for a massive scalar field $\Phi$ reads

\begin{equation}
\frac{1}{\sqrt{-G}}\partial_{\mu}\left(\sqrt{-G}\,G^{\mu\nu}\partial_{\nu}\Phi\right)-\frac{m^{2}}{\hbar^{2}}\Phi=0.
\end{equation}

We can apply the semiclassical WKB approximation for
the scalar field $\Phi$ as follows
\begin{equation}
\Phi(t,r,\theta,\phi,\psi)=\exp{\left(\frac{i}{\hbar}S\left(t,r,\theta,\phi,\psi\right)\right)}.
\end{equation}

Then we recover the following equation
\begin{equation}\label{83}
\frac{1}{F}(\partial_{t}S)^{2}=\frac{1}{H}(\partial_{r}S)^{2}+\frac{1}{K}(\partial_{\theta}S)^{2}+\frac{N}{MN-P^{2}}(\partial_{\phi}S)^{2}+\frac{M}{MN-P^{2}}(\partial_{\psi}S)^{2}-\frac{2P}{MN-P^{2}}(\partial_{\phi}S)(\partial_{\psi}S)+m^{2}.
\end{equation}

Furthermore, we can choose the same form of the action as in the last section
\begin{equation}
S_{0}(t,r,\theta,\phi,\psi)=-(E-(j\Omega_{a}+l\Omega_{b}))t+R(r,\theta)+j\phi+l\psi.
\end{equation}

Substituting this action into the Eq. \eqref{83} yields 
\begin{equation}
\frac{1}{F}(E-(j\Omega_{a}+l\Omega_{b}))^{2}=\frac{1}{H}(R^{\prime})^{2}+\frac{1}{K}(\partial_{\theta}R)^{2}+\frac{N}{MN-P^{2}}j^{2}+\frac{M}{MN-P^{2}}l^{2}-\frac{2P}{MN-P^{2}}jl+m.
\end{equation}

Solving for the radial part is not difficult to show that 
\begin{equation}
R(r)=\pm\int \frac{1}{\sqrt{\mathcal{H}^{\prime}(r_{+}) F^{\prime}(r_{+})}}\sqrt{(E-(j\Omega_{a}+l\Omega_{b}))^{2}-F(r)\left[m^{2}+\frac{1}{MN-P^{2}}\left(Nj^{2}+Ml^{2}-2Pjl\right)+\frac{(\partial_{\theta}R)^{2}}{K}\right]}\mathrm{d}r.
\end{equation}

In other words, we have shown that the same black hole temperature can be recovered in the co-rotating frame
\begin{equation}
T_{H}=\frac{\partial_{r}\Pi-2r_{0}^{2}r_{+}}{4\pi r_{0}^{2}r_{+}^{2}}.
\end{equation}

We therefore conclude that the Hawking temperature for 5D-MPBH is independent of the selected coordinate system. 

\section{Conclusion}

In this paper, for 5D-MPBH which has multi-rotation parameters, our results fill in the gap existing in the literature applying the Hamilton-Jacobi tunneling method. We have investigated the tunneling effect of the 5D-MPBH with
two independent angular momentum components using the Hamilton-Jacobi method. Furthermore we have calculated the effect of the rotation on the
Hawking radiation of scalar particles and spin-1 vector particles
from the 5D-MPBH. Firstly, we have calculated the Hawking temperature of massive vector and scalar particles from the 5D-MPBH in the Painlev\'{e} coordinates and then in the co--rotating frame by applying the WKB method and Hamilton-Jacobi equation.

The original Hawking temperature of the 5D-MPBH is impeccably obtained
in the both coordinate systems in full agreement with \cite{frolov,zheng}. Hence the main result is that the Hawking temperature is independent
of the selected coordinate system. Our future project is to investigate
the possible role of quantum horizon fluctuations on the Hawking radiation \cite{Albert}.

\section{Acknowledgements}
We thank the anonymous referees and editor for their valuable and constructive suggestions.

\end{document}